\documentclass[fleqn,12pt,twoside]{article}
\usepackage{espcrc2}

\usepackage{graphicx}
\usepackage[figuresright]{rotating}

\newcommand{\AmS}{{\protect\the\textfont2
  A\kern-.1667em\lower.5ex\hbox{M}\kern-.125emS}}

\title{What the lattice can tell us about nucleon structure}

\author{W. Schroers
\address{John von Neumann-Institut f\"ur Computing NIC/DESY, 15738 Zeuthen, Germany}}
       
\begin{document}

\begin{abstract}
This review focuses on the current status of lattice calculations of
three observables which are both phenomenologically and experimentally
relevant and have been scrutinized recently. These three observables
are the nucleon electromagnetic form factors, the momentum fraction,
$\langle x\rangle_{\mbox{\tiny u-d}}$, and the nucleon axial coupling,
$g_A$.

\hfill DESY preprint: DESY 05-243
\end{abstract}

\maketitle

\section{INTRODUCTION}
\label{sec:introduction}
Understanding nucleon structure using lattice simulations has
progressed tremendously over the past years. Both the availability of
new machines and the improvement of algorithms and techniques have
contributed to the progress. The major challenge of contemporary
lattice simulations remains to perform simulations at sufficiently
small quark masses. For example, the discovery of Ginsparg-Wilson
fermions has made it possible to calculate at quark masses which are
substantially lighter than what used to be accessible before.

On the other hand, chiral extrapolation techniques are increasingly
better understood. The current status of research depends on the
quantity under consideration. For a recent review on chiral
extrapolation techniques for nucleon structure, see
\cite{Gockeler:2004vf}. For a general review on lattice chiral
perturbation techniques, see \cite{Bar:2004xp}. An interesting new
development is also the application of models, like the chiral
quark-soliton model in \cite{Goeke:2005fs}. However, such models have
not yet been used to make contact with nucleon structure lattice data
so far.

This presentation focuses on the current status of three observables
characterizing the shape of the nucleon, the electromagnetic form
factors, $F_1$ and $F_2$, the first moment of the nucleon
spin-independent parton distribution, $\langle x\rangle_{\mbox{\tiny
u-d}}$ and the axial coupling, $g_A$. The results for $\langle
x\rangle_{\mbox{\tiny u-d}}$ will be quoted in the
$\overline{\mbox{MS}}$-scheme with a scale of $\mu=2$ GeV. The
discussion of the latter two quantities constitutes an update to the
previous review \cite{Schroers:2005rm}.

For technical details of lattice methods see
\cite{Gockeler:1995wg,Gockeler:2004wp,Hagler:2004brp} and also
\cite{Horsley:2004uq} for a recent review. For calculations of GPDs
going beyond experimentally accessible data, see references
\cite{Gockeler:2003jf,Negele:2004aa}.

\begin{table*}[htb]
\caption{Compilation of different lattice investigations of $\langle
  x\rangle_{\mbox{\tiny u-d}}$ and $g_A$.}
\label{tab:review}
\begin{tabular}[c]{@{}*{5}{l}}
\hline
Group \& Ref. & $m_\pi$ & Technique & $\langle
x\rangle_{\mbox{\tiny u-d}}$ & $g_A$ \\
\hline
Kentucky \cite{Liu:1992ab}   & ? & Wilson (quenched)             & - & $1.20(11)$ \\
KEK \cite{Fukugita:1994fh}   & $>530$ MeV & Wilson (quenched)    & - & $0.985(25)$ \\
QCDSF \cite{Gockeler:1995wg} & $>600$ MeV & Wilson (quenched) & $0.263(17)$ & $1.074(90)$ \\
LHPC  \cite{Dolgov:2002pr}   & $>650$ MeV & Wilson (full)        & $0.269(23)$ & $1.031(81)$ \\
RBCK  \cite{Sasaki:2003jh}   & $>390$ MeV & DWF (quenched)       & - & $1.212(27)$ \\
LHPC  \cite{Edwards:2005kw}  & $>360$ MeV & Hybrid & -$^{\mbox{\tiny 1}}$ & -$^{\mbox{\tiny 1}}$ \\
LHPC  \cite{Edwards:2005ym}  & $>360$ MeV & Hybrid & - & $1.212(84)$ \\
QCDSF \cite{Gockeler:2004wp} & $>550$ MeV & CI-Wilson (quenched) & $0.245(19)$ & - \\
QCDSF \cite{Khan:2004vw}     & $>550$ MeV & CI-Wilson (full)     & - & -$^{\mbox{\tiny 1}}$ \\
QCDSF \cite{Khan:2005ns}     & $>550$ MeV & CI-Wilson (full)     & - & -$^{\mbox{\tiny 2}}$ \\
QCDSF \cite{Galletly:2005db} & $>300$ MeV & Overlap (quenched)   & -$^{\mbox{\tiny 1}}$ & -$^{\mbox{\tiny 1}}$ \\
RBCK  \cite{Ohta:2004mg,Orginos:2005uy}
                             & $>400$ MeV & DWF (quenched/full)  & -$^{\mbox{\tiny 3}}$ & - \\
\hline
\multicolumn{5}{l}{Experimental values: $\langle x\rangle_{\mbox{\tiny
      u-d}}=0.154(3)$ \cite{Lai:1996mg}, $g_A=1.248(2)$
  \cite{Blumlein:2002be}, see also
  \cite{Blumlein:2004ip}} \\ \hline
\end{tabular}\\[2pt]
$^{\mbox{\tiny 1}}$ Work in progress and/or no prediction quoted \\
$^{\mbox{\tiny 2}}$ Constrained fit yields consistency with experiment
\\ $^{\mbox{\tiny 3}}$ Agreement found with experiment for $\langle
x\rangle_{\mbox{\tiny u-d}}/\langle x\rangle_{\mbox{\tiny $\Delta$
u-$\Delta$ d}}$
\end{table*}

Results from different lattice studies are shown in
table~\ref{tab:review}. These studies combine a variety of different
technologies and cover a wide range of parameters. While the early
investigations were limited to linear chiral extrapolations from
rather large quark masses, the latter studies either do not quote
numbers from naive extrapolations or perform constrained fits using
chiral perturbation theory.

\section{ELECTROMAGNETIC FORM FACTORS}
\label{sec:elmag-ff}
Understanding the behavior of the form factor ratio, $F_2/F_1$, has
recently regained attention in the study of exclusive
reactions. Arguments based on asymptotic scaling \cite{Brodsky:1974vy}
result in scaling laws for $F_1\propto Q^{-4}$ and $F_2\propto
Q^{-6}$. Hence, one would expect $F_2/F_1$ to scale like $\propto
Q^{-2}$. Recent experiments \cite{Gayou:2001qd}, however, have
indicated that instead the data is more consistently described with a
scaling relation like $F_2/F_1\propto Q^{-1}$. Using arguments from
perturbative QCD, reference \cite{Belitsky:2002kj} finds an explicit
expression for the scaling behavior of the form factor ratio which is
consistent with the experimental data.

First lattice studies have confirmed this scaling behavior already at
quark masses beyond $700$ MeV \cite{Negele:2004aa}. So far, the QCDSF
group is currently investigating this ratio in full QCD using Clover
fermions \cite{Pleiter:2005ta,Gockeler:2005pr}. An obstacle is still
provided by the fact that lattice data is only available for values of
$Q^2$ smaller than $3$ GeV$^2$. A preliminary feasibility study of the
behavior of form factors at larger values of $Q^2$ has been performed
in \cite{Edwards:2005kw}. It appears that the region of $Q^2>3$
GeV$^2$ is currently not easily reached in lattice simulations.

Beyond this ratio, the computation of magnetic moments and rms-radii
is possible using lattice simulations today. Unlike the asymptotic
ratio these quantities require data points at small values of $Q^2$
with sufficient accuracy. In principle, the quality of the lattice
data is almost as good as for $g_A$ since the matrix elements do not
involve any derivatives. However, at the moment there are only few
sophisticated expressions from small quark mass expansions known. The
direct comparison between lattice data with one variant of a small
scale expansion at values of the pion mass beyond $550$ MeV has been
presented in \cite{Gockeler:2003ay}. The lattice data in this
publication, however, is quenched and it will be interesting to
perform a careful check with unquenched data which will soon be
available with good statistics in \cite{Gockeler:2005pr}. For the case
of magnetic moments, a quenched study in \cite{Young:2004tb} has
attempted a comparison between quenched and real QCD beyond the
application of quenched chiral effective field theory. A different
path to compute magnetic moments has been pursued in
\cite{Lee:2005ds}.

It appears that still some technological improvements and an increased
understanding of chiral techniques is necessary to finally arrive at
quantitative predictions similar to those available for $g_A$.

\section{THE MOMENT $\langle x\rangle_{\mbox{\tiny u-d}}$}
\label{sec:moment-x}
As has been pointed out before in \cite{Schroers:2005rm}, almost all
results for $\langle x\rangle_{\mbox{\tiny u-d}}$ systematically
exceed the results quoted by phenomenologists by about $50\%$. Given
the variety of parameters and technologies used, neither finite size,
nor unquenching or lattice artifacts are responsible for this
discrepancy.

While an earlier study by the QCDSF collaboration using quenched
Overlap fermions in \cite{Gurtler:2004ac} found a systematically
smaller value than other publications, an update of this calculation
in \cite{Galletly:2005db} finds results consistent with those of
Wilson-type fermions. This discrepancy has been resolved in
\cite{Gurtler:2005ym}. It was found that the discrepancy can arise
from considering a non-perturbative matching instead of leading-order
perturbative renormalization.

A mismatch similar to the one found in \cite{Gurtler:2004ac} has been
reported by the LHPC collaboration in \cite{Edwards:2005kw}, see
figure 3 in this reference. The data points for $\langle
x\rangle_{\mbox{\tiny u-d}}$ lie on a straight line with no indication
of any bending down. However, they are both inconsistent with the
results at large quark masses and the experimental value. This paper
also utilizes a leading order perturbative matching and it is possible
that a non-perturbative calculation of the renormalization procedure
will lead to a correction that makes the hybrid data consistent with
the other calculations again.

It is interesting as the RBCK collaboration points out in
\cite{Orginos:2005uy} that the ratio $\langle x\rangle_{\mbox{\tiny
u-d}}/\langle x\rangle_{\mbox{\tiny $\Delta$ u-$\Delta$ d}}$ is in
complete agreement with the experimental value. This finding has also
been reproduced by LHPC in \cite{Edwards:2005kw}. It could provide an
indication that the problem affects both individual quantities in the
same way by a common factor which cancels as the ratio is taken.

At this moment it is safe to conclude that no study so far has
observed any evidence for a ``bending down'' of the momentum fraction
as the quark mass is decreased. A possible scenario for such a
behavior has been explored in \cite{Detmold:2001jb}. The mystery
shrouding this quantity remains unsolved.

\section{THE AXIAL COUPLING $g_A$}
\label{sec:gA}
Unlike the observable discussed in the previous section, the
understanding of the axial coupling has progressed much
further. Although it is not yet possible to postdict the experimental
value of $g_A$ from first principles alone, it has become feasible to
connect expressions from chiral expansion techniques with lattice
results.

The coupling $g_A$ is known to be very sensitive to finite-size
effects which has first been observed in \cite{Sasaki:2003jh} and
later confirmed by other groups in \cite{Khan:2004vw} and
\cite{Renner:2004ck}. Together with the dependence on the quark mass
these finite size effects can accurately be modeled by chiral
expansion techniques. These have now reached very sophisticated levels
\cite{Khan:2004vw,Beane:2004ch,Khan:2005ns}.

The reference \cite{Edwards:2005ym} reports a statistical residual
error of $5\%$ for a complete calculation of $g_A$ down to pion masses
of $350$ MeV with a physical lattice size of $3.5$ fm. This lattice
size results in negligible finite-size effects. Performing a
constrained three-parameter fit with an expression from chiral
perturbation theory finally yields an error band of $7\%$ at the
physical value of the quark mass, cf.~figure 2 in
\cite{Edwards:2005ym}.

This analysis shows how lattice simulations can already today
reproduce experimental values and gives rise to the hope that similar
quantities will soon be equally well understood.

\section{CONCLUSIONS}
\label{sec:conclusions}
The dependence of the first moment of the spin-independent parton
distribution $\langle x\rangle_{\mbox{\tiny u-d}}$ is still not
understood and there is evidence that a ``bending over'' of the
lattice data does not occur until very small values of the quark mass
if ever. This situation remains a puzzle to both phenomenologists and
lattice physicists.

However, lattice calculations are now reaching a turning point. For
the first time, it is possible to present a convincing case for the
matching of state-of-the-art lattice data and sophisticated chiral
expansion techniques. This has allowed to qualitatively and
quantitatively understand the quark mass dependence of the nucleon
axial coupling $g_A$.

A similar level can be expected to be reached for form factors in the
region of small values of $Q^2$. This will allow for an accurate
quantitative computation of rms-radii and magnetic moments. The
scaling behavior of form factors at larger values of $Q^2$, however,
appears to be a bigger obstacle and does not promise to be understood
similarly well in the near future.

This presentation was supported in part by the DFG (Forschergruppe
Gitter-Hadronen-Ph{\"a}nomenologie and in part by the EU Integrated
Infrastructure Initiative Hadron Physics (I3HP) under contract number
RII3-CT-2004-506078. I thank the Alexander-von-Humboldt Foundation for
their support and the Center for Theoretical Physics at MIT for their
hospitality.


\begin{thebibliography}{100}
\bibitem{Gockeler:2004vf} M.~G{\"o}ckeler, arXiv:hep-lat/0412013.
\bibitem{Bar:2004xp} O.~B{\"a}r, Nucl.\ Phys.\ Proc.\ Suppl.\ {\bf
  140}, 106 (2005).
\bibitem{Goeke:2005fs} K.~Goeke, J.~Ossmann, P.~Schweitzer and
  A.~Silva, arXiv:hep-lat/0505010.
\bibitem{Schroers:2005rm} W.~Schroers, Nucl.\ Phys.\ A {\bf 755}
  (2005) 333.
\bibitem{Gockeler:1995wg} M.~G{\"o}ckeler {\it et al.} [QCDSF], Phys.\ 
  Rev.\ D {\bf 53} (1996) 2317.
\bibitem{Gockeler:2004wp}  M.~G{\"o}ckeler, R.~Horsley, D.~Pleiter, P.E.L.~Rakow and
  G.~Schierholz [QCDSF Collaboration],
  arXiv:hep-ph/0410187.
\bibitem{Hagler:2004brp} M.~G{\"o}ckeler {\it et al.}  [QCDSF
  Collaboration], Nucl.\ Phys.\ A {\bf 755} (2005) 537.
\bibitem{Horsley:2004uq} R.~Horsley, arXiv:hep-lat/0412007.
\bibitem{Gockeler:2003jf} M.~G{\"o}ckeler {\it et al.} [QCDSF], Phys.\ 
  Rev.\ Lett.\ {\bf 92}, 042002 (2004). \\
  P.~H{\"a}gler {\it et al.} [LHPC], Phys.\ Rev.\ D {\bf 68}, 034505
  (2003). \\
  P.~H{\"a}gler {\it et al.} [LHPC], Phys.\ Rev.\ Lett.\ {\bf 93},
  112001 (2004). \\
  J.~W.~Negele {\it et al.}  [LHPC], Nucl.\ Phys.\ Proc.\ 
  Suppl.\ {\bf 129}, 910 (2004). \\
  W.~Schroers {\it et al.}  [LHPC], Nucl.\ Phys.\ Proc.\ 
  Suppl.\ {\bf 129}, 907 (2004). \\
  M.~G{\"o}ckeler {\it et al.}  [QCDSF], Nucl.\ Phys.\ 
  Proc.\ Suppl.\ {\bf 128}, 203 (2004). \\
  M.~G{\"o}ckeler {\it et al.}  [QCDSF], Nucl.\ Phys.\ 
  Proc.\ Suppl.\ {\bf 135}, 156 (2004). \\
  M.~G{\"o}ckeler {\it et al.}  [QCDSF Collaboration], Nucl.\ Phys.\
  Proc.\ Suppl.\ {\bf 140} (2005) 399. \\
  P.~H{\"a}gler, J.~W.~Negele, D.~B.~Renner, W.~Schroers, T.~Lippert
  and K.~Schilling [LHPC Collaboration], Eur.\ Phys.\ J.\ A {\bf 24S1}
  (2005) 29. \\
  M.~G{\"o}ckeler {\it et al.}  [QCDSF Collaboration], Few Body Syst.\
  {\bf 36} (2005) 111. \\
  B.~Bistrovic {\it et al.}  [Lattice Hadron Physics Collaboration],
  J.\ Phys.\ Conf.\ Ser.\ {\bf 16} (2005) 150.
\bibitem{Negele:2004aa} J.~W.~Negele {\it et al.} [LHPC], Nucl.\
    Phys.\ Proc.\ Suppl.\ {\bf 128}, 170 (2004).
\bibitem{Liu:1992ab} K.~F.~Liu {\it et al.}, Phys.\ Rev.\ D {\bf 49},
  4755 (1994).
\bibitem{Fukugita:1994fh} M.~Fukugita, Y.~Kuramashi, M.~Okawa and
  A.~Ukawa, Phys.\ Rev.\ Lett.\ {\bf 75} (1995) 2092.
\bibitem{Dolgov:2002pr} D.~Dolgov {\it et al.} [LHPC], Phys.\ Rev.\ D
  {\bf 66} (2002) 034506.
\bibitem{Sasaki:2003jh} S.~Sasaki, K.~Orginos, S.~Ohta and T.~Blum
  [RBCK], Phys.\ Rev.\ D {\bf 68}, 054509 (2003).
\bibitem{Edwards:2005kw} R.~G.~Edwards {\it et al.} [LHPC
  Collaboration], PoS {\bf LAT2005} (2005) 056.
\bibitem{Edwards:2005ym} R.~G.~Edwards {\it et al.}  [LHPC
  Collaboration], arXiv:hep-lat/0510062.
\bibitem{Khan:2004vw} A.~A.~Khan {\it et al.}, Nucl.\ Phys.\ Proc.\
  Suppl.\  {\bf 140}, 408 (2005).
\bibitem{Khan:2005ns} A.~A.~Khan {\it et al.}, arXiv:hep-lat/0510061.
\bibitem{Galletly:2005db} D.~Galletly {\it et al.}  [QCDSF
  Collaboration], arXiv:hep-lat/0510050.
\bibitem{Ohta:2004mg} S.~Ohta and K.~Orginos [RBCK], Nucl.\ Phys.\
  Proc.\ Suppl.\  {\bf 140}, 396 (2005). \\
  S.~Ohta and K.~Orginos [RBCK], Nucl.\ Phys.\ Proc.\ 
  Suppl.\ {\bf 129} (2004) 296.
\bibitem{Orginos:2005uy} K.~Orginos, T.~Blum and S.~Ohta,
  arXiv:hep-lat/0505024.
\bibitem{Lai:1996mg} H.~L.~Lai {\it et al.}, Phys.\ Rev.\ D {\bf 55},
  1280 (1997). \\
  M.~Gl{\"u}ck, E.~Reya and A.~Vogt, Eur.\ Phys.\ J.\ C {\bf 5}, 461
  (1998). \\
  A.~D.~Martin, R.~G.~Roberts, W.~J.~Stirling and R.~S.~Thorne, Eur.\ 
  Phys.\ J.\ C {\bf 23}, 73 (2002).
\bibitem{Blumlein:2002be} J.~Bl{\"u}mlein and H.~B{\"o}ttcher, Nucl.\ 
  Phys.\ B {\bf 636} (2002) 225. \\
  M.~Gl{\"u}ck, E.~Reya, M.~Stratmann and W.~Vogelsang, Phys.\ Rev.\ D
  {\bf 63} (2001) 094005. \\
  Y.~Goto {\it et al.} [AAC], Phys.\ Rev.\ D {\bf 62},
  034017 (2000). \\
  M.~Hirai, AIP Conf.\ Proc.\ {\bf 675}, 365 (2003).
\bibitem{Blumlein:2004ip} J.~Bl{\"u}mlein, H.~B{\"o}ttcher and
  A.~Guffanti, Nucl.\ Phys.\ Proc.\ Suppl.\ {\bf 135} (2004) 152.
\bibitem{Brodsky:1974vy} S.~J.~Brodsky and G.~R.~Farrar, Phys.\ Rev.\
  D {\bf 11} (1975) 1309. \\
  G.~P.~Lepage and S.~J.~Brodsky, Phys.\ Rev.\ D {\bf 22} (1980) 2157.
\bibitem{Gayou:2001qd} O.~Gayou {\it et al.}  [Jefferson Lab Hall A
  Collaboration], Phys.\ Rev.\ Lett.\ {\bf 88} (2002) 092301. \\
  O.~Gayou {\it et al.}, Phys.\ Rev.\ C {\bf 64} (2001) 038202.
\bibitem{Belitsky:2002kj} A.V.~Belitsky, X.~d.~Ji and F.~Yuan, Phys.\
  Rev.\ Lett.\ {\bf 91}, 092003 (2003).
\bibitem{Pleiter:2005ta} D.~Pleiter, ``Nucleon form factors from
  $n_f=2$ clover fermions'', talk presented at the ILFT workshop at
  Jefferson Lab, Oct.~2005.
\bibitem{Gockeler:2005pr} M.~G{\"o}ckeler {\it et al.}, in
preparation.
\bibitem{Gockeler:2003ay} M.~G{\"o}ckeler, T.R.~Hemmert, R.~Horsley,
 D.~Pleiter, P.E.L.~Rakow, A.~Sch{\"a}fer and G.~Schierholz [QCDSF
 Collaboration], Phys.\ Rev.\ D {\bf 71} (2005) 034508.
\bibitem{Young:2004tb} R.D.~Young, D.B.~Leinweber and A.W.~Thomas,
  Phys.\ Rev.\ D {\bf 71} (2005) 014001.
\bibitem{Lee:2005ds} F.~X.~Lee, R.~Kelly, L.~Zhou and W.~Wilcox,
  Phys.\ Lett.\ B {\bf 627}, 71 (2005).
\bibitem{Gurtler:2004ac} M.~G{\"u}rtler {\it et al.},
  arXiv:hep-lat/0409164. \\
  D.~Galletly {\it et al.} [QCDSF-UKQCD], Nucl.\ Phys.\ 
  Proc.\ Suppl.\ {\bf 129}, 453 (2004).
\bibitem{Gurtler:2005ym} M.~G{\"u}rtler, R.~Horsley, P.~E.~L.~Rakow,
  C.~J.~Roberts, G.~Schierholz and T.~Streuer [QCDSF Collaboration],
  arXiv:hep-lat/0510045.
\bibitem{Detmold:2001jb} W.~Detmold {\it et al.}, Phys.\ Rev.\ Lett.\ 
  {\bf 87}, 172001 (2001).
\bibitem{Renner:2004ck} D.~B.~Renner {\it et al.} [LHPC], Nucl.\
  Phys.\ Proc.\ Suppl.\  {\bf 140}, 255 (2005).
\bibitem{Beane:2004ch} S.R.~Beane and M.J.~Savage, Phys.\ Rev.\ D {\bf
  70} (2004) 074029. \\
  W.~Detmold and D.J.D.~Lin, Phys.\ Rev.\ D {\bf 71} (2005) 054510.
\end{thebibliography}
\end{document}